\documentclass[12pt]{iopart}
\usepackage{iopams}
\usepackage{graphicx}
\usepackage[T1]{fontenc}
\usepackage{ae}
\usepackage[latin1]{inputenc}
\usepackage{color}

\newcommand{\Eins}
           {\;\smash{\raisebox{-0.5ex}{$\!\!\stackrel{\!\mbox{1}
            \hspace{-0.4ex}\rule[0.0ex]{0.06ex}{1.60ex}}{ }$}}}
\newcommand{\op}[1]{%
    \fontdimen12\textfont3=2pt\fontdimen12\scriptfont3=1.4pt%
    \!\null\mathop{\vphantom{#1}\smash{#1}}\limits_{\sim}\null\!}

\newcommand{\binom}[2]{{#1 \choose #2}}

\newcommand{\di}{\unitlength0.5cm\begin{picture}(1,1)
\put(0.,0.2){\circle*{0.2}}
\put(0.5,0.2){\circle*{0.2}}
\put(0,0.2){\line(1,0){0.5}}
\put(0.1,0.4){\footnotesize $1$}
\end{picture}}

\newcommand{\tii}{\unitlength0.6cm\begin{picture}(1,1)
\put(0.,0.2){\circle*{0.2}}
\put(0.5,0.2){\circle*{0.2}}
\put(1,0.2){\circle*{0.2}}
\put(0,0.2){\line(1,0){0.5}}
\put(0.5,0.2){\line(1,0){0.5}}
\put(0.1,0.4){\footnotesize $k_1$}
\put(0.6,0.4){\footnotesize $k_2$}
\end{picture}}

\newcommand{\ti}{\unitlength0.6cm\begin{picture}(1,1)
\put(0.,0.2){\circle*{0.2}}
\put(0.5,0.2){\circle*{0.2}}
\put(1,0.2){\circle*{0.2}}
\put(0,0.2){\line(1,0){0.5}}
\put(0.5,0.2){\line(1,0){0.5}}
\put(0.1,0.4){\footnotesize $1$}
\put(0.6,0.4){\footnotesize $k_2$}
\end{picture}}

\newcommand{\dii}{\unitlength0.5cm\begin{picture}(1,1)
\put(0.,0.2){\circle*{0.2}}
\put(0.5,0.2){\circle*{0.2}}
\put(0,0.2){\line(1,0){0.5}}
\put(0.1,0.4){\footnotesize $k$}
\end{picture}}

\begin{document}
%\today

\title[The general spin triangle]
{The general spin triangle}

\author{Heinz-J\"urgen Schmidt$^1$
 \footnote[3]{Correspondence should be addressed to
hschmidt@uos.de}  }

\address{$^1$ Universit\"at Osnabr\"uck, Fachbereich Physik,
Barbarastr. 7, D - 49069 Osnabr\"uck, Germany}

\begin{abstract}
We consider the Heisenberg spin triangle with general coupling coefficients and general spin quantum number $s$.
The corresponding classical system is completely integrable.
In the quantum case the eigenvalue problem can be reduced to that of tridiagonal matrices in at most $2s+1$ dimensions.
The corresponding energy spectrum exhibits what we will call spectral symmetries due to the underlying permutational
symmetry of the considered class of Hamiltonians. As an application we explicitly calculate six classes
of universal polynomials that occur in the high temperature expansion of spin triangles and more general spin systems.
Aspects of quantum integrability are also discussed in this context.
\end{abstract}

%Uncomment for PACS numbers title message
%\pacs{75.10.b, 75.10.Jm}

% Uncomment for Submitted to journal title message
%\submitto{\JPA}

% Comment out if separate title page not required
\maketitle

\section{Introduction}
\label{sec:I}
%%%%%%%%%%%%%%%%%%%%%%%%%%%%%%%%%%%%%%%%%%%%%%%%%%%%%%%%%%%%%%%%%%%%%%%%%%%%%%%%%%%%%%%%%%%%%%%%%%%%%%%%%%%%%%%%%%%%%%%%%%%%%%%%%%%%%%%%%%%%%%%%%%%%%
Theoretical investigations of spin systems are often motivated by the
applications to real systems as magnetic molecules or spin lattices.
In the case of a system of three spins with realistic spin quantum numbers varying from
$s=\frac{1}{2}$ to, say, $s=\frac{7}{2}$, the underlying Hilbert space has a maximal dimension
of $(2s+1)^3 = 8^3=512$. Hence, even without using any dimensional reduction due to rotational symmetry,
it is an easy exercise to numerically calculate physically relevant quantities, e.~g.~, specific heat
or magnetization curves, by using standard codes. Other aspects of spin triangles have been treated
in recent publications \cite{AK}, \cite{HBM}, and \cite{SFA}.
Thus a detailed theoretical investigation
of the general spin triangle requires additional justification.
In my opinion, the reasons to study this special class of spin systems are threefold.\\

First, the spin triangle is interesting in its own right since it exhibits salient features
which might evade the purely numerical treatment. It is a straight forward task to reduce the
pertinent Hamiltonian $H$ to that of a two-chain with coupling constants $1$ and $y$ and to restrict
it to invariant subspaces of maximal dimension $2s+1$, see section \ref{sec:D}. Here the reduced
Hamiltonian has a tridiagonal matrix representation, see section \ref{sec:H}, which can be understood
by virtue of the Wigner-Eckhard theorem. Some properties of tridiagonal matrices that are relevant for our purpose are
collected in Appendix A. The corresponding characteristic polynomial and the spectral
curves $e_\nu(y)$ exhibit a number of symmetries that I have dubbed ``spectral symmetries" in section \ref{sec:S}.
Usually, symmetries in quantum mechanics lead to operators commuting with $H$. Spectral symmetries are
of a different kind and can be expressed by functional equations for the $e_\nu(y)$. They can be traced back
to permutational invariance, not of a particular Hamiltonian, but of the whole class of triangle Hamiltonians.
Despite its elementary origin, spectral symmetry allows a qualitative discussion of the spectral curves and, e.~g.~,
the calculation of the moments of the restricted Hamiltonian
up to fifth order without using its explicit matrix representation.\\

A second reason to study spin triangles is that their properties might help to understand
more complex systems. One example, not considered in this article, is the $S=0$ ground state
(for integer $s$ and approximately equal coupling constants) that has been used  to build ground
states of larger spin systems including spin lattices, see \cite{SR10} and \cite{RS11}.
Another example is the observation that the first few terms of the perturbation series
for the eigenvalues of $H$ yield certain classes of universal polynomials that occur
in the graph expansion of the $k$-th moments of $H$, see section \ref{sec:M} and \ref{sec:T}.
These polynomials also occur in the high temperature expansion for general Heisenberg systems,
not only triangles, see \cite{RBW} and \cite{SLR}. Appendix B contains explicit expressions
for the $k$-th moments of spin dimers that are used in the formulas for the the above-mentioned polynomials.\\

Finally, spin triangles may serve as simple but non-trivial examples to study theoretical problems of general interest.
Here we will concentrate on one aspect, namely integrability. Classical systems are completely integrable
in the sense of Arnold-Liouville iff they possess $N$ Poisson-commuting integrals of motion, where $2N$ is
the phase space dimension. Sometimes the corresponding quantum systems are also analytically solvable
(harmonic oscillator, hydrogen atom) but not always. One counter-example is the one-dimensional particle in a
potential $V(x)$, another one is the general spin triangle. In the latter case we have classically three unit spin
vectors, i.~e.~, a six-dimensional phase space, and three constants of motion, $H, \, S^2$ and $S_3$. The corresponding
quantum system, notwithstanding the partial results of this paper, cannot be solved in closed form. However,
there are weaker notions of ``quantum integrability" that justify, e.~g.~, to consider the $s=\frac{1}{2}$ chain
as integrable by means of the Bethe ansatz , see \cite{Bethe} or \cite{F} for generalizations.
In this context one could see the results of the present paper as another example
within the wide range of the not yet fully understood notion of quantum integrability.
We will come back to this question in
the section \ref{sec:SO} containing the summary and an outlook.

%%%%%%%%%%%%%%%%%%%%%%%%%%%%%%%%%%%%%%%%%%%%%%%%%%%%%%%%%%%%%%%%%%%%%%%%%%%%%%%%%%%%%%%%%%%%%%%%%%%%%%%%%%%%%%%%%%%%%%%%%%%%%%%%%%%%%%%%%%%%%%%%

\section{Basic definitions and elementary results}\label{sec:D}

%%%%%%%%%%%%%%%%%%%%%%%%%%%%%%%%%%%%%%%%%%%%%%%%%%%%%%%%%%%%%%%%%%%%%%%%%%%%%%%%%%%%%%%%%%%%%%%%%%%%%%%%%%%%%%%%%%%%%%%%%%%%%%%%%%%%%%%%%%%%%%%%

\begin{center}
\begin{figure}
\begin{center}
\includegraphics[width=8cm]{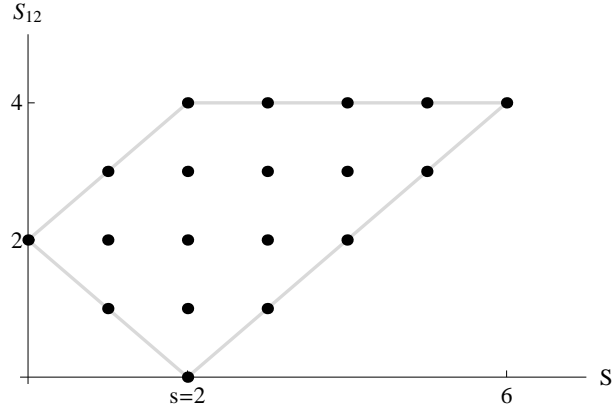}
\end{center}
\caption{\label{figG1}The range of the quantum numbers $S_{12}$ and
$S$ for $s=2$. It is typical for integer $s$ that $S$ has a minimal
value of $S_{\mbox{\footnotesize min}}=0$ and the corresponding
eigenspace $\mathcal{H}(S)$ will be one-dimensional. Note that
dimensions of the various spaces $\mathcal{H}(S)$ are
$1+3+5+4+3+2+1=19=3s(s+1)+1$ in accordance with (\ref{D11}).}
\end{figure}
\end{center}

\begin{center}
\begin{figure}
\begin{center}
\includegraphics[width=8cm]{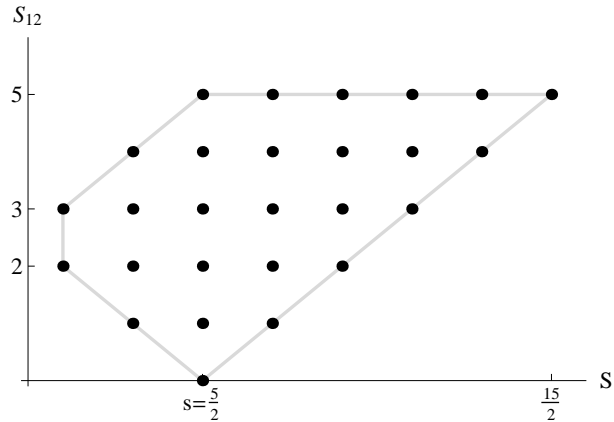}
\end{center}
\caption{\label{figG2}The range of the quantum numbers $S_{12}$ and
$S$ for $s=5/2$. It is typical for half-integer $s$ that $S$ has a
minimal value of $S_{\mbox{\footnotesize min}}=1/2$ and the
corresponding eigenspace $\mathcal{H}(S)$ will be two-dimensional.
Note that dimensions of the various spaces $\mathcal{H}(S)$ are
$2+4+6+5+4+3+2+1=27=3s(s+1)+3/4$ in accordance with (\ref{D11}).}
\end{figure}
\end{center}

We consider the general Heisenberg spin triangle with one and the
same individual spin quantum number $s=\frac{1}{2},1,\frac{3}{2},2,\ldots$ and the
Hamiltonian
\begin{equation} \label{D1}
\op{H}\,
=
\,J_1\op{\bi{s}}_2\cdot\op{\bi{s}}_3\,
+
\,J_2\op{\bi{s}}_3\cdot\op{\bi{s}}_1\,
+
\,J_3\op{\bi{s}}_1\cdot \op{\bi{s}}_2\,
\end{equation}
where
the $J_1,\,J_2,\,J_3$ are three arbitrary real parameters and the $\op{\bi{s}}_i$
denote the three individual spin vector operators, $i=1,2,3$. The Hamiltonian (\ref{D1})
can be analytically diagonalized if two of the $J_i$'s coincide (i.~e.~for the isosceles triangle)
or if $s$ is small in a sense to be made more precise subsequently.
In the first case of, say, $J_1=J_2\equiv J$
we have
\begin{eqnarray} \label{D2a}
\op{H}\,=\,\op{H}_J
&=&
\,J\,(\op{\bi{s}}_1+\op{\bi{s}}_2)\cdot\op{\bi{s}}_3\,
+
\,J_3\,\op{\bi{s}}_1\cdot\op{\bi{s}}_2\\
\label{D2b}
&=&
\frac{J}{2}\,\op{\bi{S}}^2 +\frac{J_3-J}{2}\op{\bi{S}}_{12}^2-
\left( \frac{J}{2}+J_3
\right)\,s(s+1)
\;,
\end{eqnarray}
where
\begin{eqnarray} \label{D3a}
\op{\bi{S}}\,
&\equiv&
\op{\bi{s}}_1+\op{\bi{s}}_2+\op{\bi{s}}_3\;,
\\
\label{D3b}
\op{\bi{S}}_{12}\,
&\equiv&
\op{\bi{s}}_1+\op{\bi{s}}_2
\;.
\end{eqnarray}
It is well-known that $\op{\bi{S}}^2$ and $\op{\bi{S}}_{12}^2$
commute and have simultaneous eigenvalues
of the form $S(S+1),\;S_{12}(S_{12}+1)$ where
\begin{eqnarray} \label{D4a}
S&=& 3s,3s-1,\ldots\,S_{\mbox{\footnotesize min}}\ge\,0
\\
\label{D4b}
S_{12}&=& 2s,2s-1,\ldots ,0
\;,
\end{eqnarray}
and the triangle inequality
\begin{equation}\label{D5}
|S_{12} -s|\le S\le S_{12}+s
\end{equation}
holds. From this the eigenvalues of $\op{H}_J$ immediately follow  from (\ref{D2b}).
The corresponding eigenbasis of $\op{H}_J$ will  be used to split
the matrix representation of the general $\op{H}$ into smaller block matrices.\\
In the general case we write
\begin{equation}\label{D6}
\op{H}=\frac{J_2}{2}
\left(
\op{\bi{S}}^2-3s(s+1)
\right)
+
(J_3-J_2)
\left(
\op{\bi{s}}_1\cdot\op{\bi{s}}_2+\frac{J_1-J_2}{J_3-J_2}
\op{\bi{s}}_2\cdot\op{\bi{s}}_3
\right)
\;,
\end{equation}
and, since $\op{H}$ and $\op{\bi{S}}^2$ commute, the problem of
diagonalizing $\op{H}$ can be reduced to the task of finding the
eigenvalues and eigenvectors of
\begin{eqnarray}\label{D7a}
\op{H}_2(y)&\equiv&
\op{\bi{s}}_1\cdot\op{\bi{s}}_2+y\,\op{\bi{s}}_2\cdot\op{\bi{s}}_3,\quad\mbox{where}\\
\label{D7b}
y&\equiv&\frac{J_1-J_2}{J_3-J_2}
\end{eqnarray}
in the invariant subspace $\mathcal{H}(S,S_3)$. The latter
denotes the common eigenspace of $\op{\bi{S}}^2$ and $\op{{S}}_3$ corresponding to
the eigenvalues $S(S+1)$ and, say,
\begin{equation}\label{D8}
S_3=S_{\mbox{\footnotesize min}}\equiv
\left\{
\begin{array}{rcl}
0&:&s\mbox{ integer}\\
\frac{1}{2}&:& s\mbox{ half-integer}
\end{array}
\right.
\end{equation}
It will be simply denoted by $\mathcal{H}(S)$, since the eigenvalues of $\op{H}_2(y)$
do not depend on the chosen eigenspace of $\op{{S}}_3$. The dimension of $\mathcal{H}(S)$
can be determined by counting the possible values of the quantum number $S_{12}$.
We obtain
\begin{eqnarray}\label{D9a}
s-S \le S_{12} \le S+s &\quad\mbox{ for }\quad & 0\le S\le s,
\\
\label{D9b}
S-s \le S_{12} \le 2s &\quad\mbox{ for }\quad & s\le S\le 3s,
\end{eqnarray}
according to the triangle inequality (\ref{D5}). In order to avoid the case
distinction (\ref{D9a}),(\ref{D9b}) in what follows, we will write the above
condition in a uniform way as
\begin{equation}\label{D10}
g_1\,\le \,g\,\le \,g_2
\;,
\end{equation}
where $g\equiv S_{12}$ and $g_1(s,S)$, $g_2(s,S)$ are defined according
to (\ref{D9a}) and (\ref{D9b}).\\
The dimension $N$ of $\mathcal{H}(S)$ increases in steps of two in
the range $0\le S\le s$ and decreases in steps of one in the range
$s\le S \le 3s$. The maximal dimension $N_{\mbox{\footnotesize
max}}=2s+1$ is assumed for $S=s$. The figures \ref{figG1} and
\ref{figG2} show typical examples. Especially, it follows that for
$s\le 3/2$ the dimension of $\mathcal{H}(S)$ will be at most four
and hence $\op{H}$ can be analytically diagonalized. At this point
one should mention that for moderate $s$, also exceeding $s=3/2$, the
computer-algebra system MATHEMATICA renders the eigenvalues of
$\op{H}_2(y)$ in the form of root objects which have many properties
of special functions. For example, the $n$-th eigenvalue of
$\op{H}_2(y)$ can be expanded into an exact power
series w.~r.~t.~$y$ up to some finite order.\\
The total dimension $d_{\mbox{\footnotesize tot}}$ of the eigenspace
of $\op{{S}}_3$ with eigenvalue $S_{\mbox{\footnotesize min}}$
amounts to
\begin{equation}\label{D11}
d_{\mbox{\footnotesize tot}}=3s(s+1)+
\left\{
\begin{array}{rcl}
1&:&s\mbox{ integer}\\
\frac{3}{4}&:& s\mbox{ half-integer}
\end{array}
\right.
\;,
\end{equation}
in accordance with the general formula in \cite{BSS}.\\

%%%%%%%%%%%%%%%%%%%%%%%%%%%%%%%%%%%%%%%%%%%%%%%%%%%%%%%%%%%%%%%%%%%%%%%%%%%%%%%%%%%%%%%%%%%%%%%%%%%%%%%%%%%%%%%%%%%%%%%%%%%%%%%%%%%%%%%%%%%%%%%%

\section{Matrix representation of the reduced Hamiltonian\label{sec:H}}

%%%%%%%%%%%%%%%%%%%%%%%%%%%%%%%%%%%%%%%%%%%%%%%%%%%%%%%%%%%%%%%%%%%%%%%%%%%%%%%%%%%%%%%%%%%%%%%%%%%%%%%%%%%%%%%%%%%%%%%%%%%%%%%%%%%%%%%%%%%%%%%%

In order to find the matrix representation of $\op{H}_2(y)$ within the invariant
subspace $\mathcal{H}(S)$ one has to choose an appropriate basis. An obvious choice
is the eigenbasis of $\op{H}_2(0)=\op{\bi{s}}_1\cdot\op{\bi{s}}_2=\frac{1}{2}
\left(\op{\bi{S}}_{12}^2-2s(s+1)\right)$. The transformation to the eigenbasis of
$\op{\bi{s}}_2\cdot\op{\bi{s}}_3$ within $\mathcal{H}(S)$
is achieved by the unitary, real, symmetric matrix $U$ with the entries
\begin{equation}\label{H1}
U_{g\,g'}=(-1)^{3s+S}\sqrt{(2g+1)(2g'+1)}
\left(
\begin{array}{ccc}
s&s&g\\
s&S&g'
\end{array}
\right)
\;,\;g_1\le g,g' \le g_2\;,
\end{equation}
where $(\ldots)$ denotes the $6J-$symbol, see, e.~g.~,
\cite{QTAM}. For the sake of simplicity we will not distinguish
between operators and their matrix representation in
$\mathcal{H}(S)$, if no misunderstanding is likely to arise. Hence we may
write
\begin{equation}\label{H2}
\op{\bi{s}}_1\cdot\op{\bi{s}}_2
=
\mbox{diag}\{\frac{1}{2}(g(g+1)-2s(s+1))| g=g_1,\ldots,g_2\}
\end{equation}
and
\begin{equation}\label{H2}
\op{\bi{s}}_2\cdot\op{\bi{s}}_3
=
U^\ast\;\op{\bi{s}}_1\cdot\op{\bi{s}}_2\;U
\;.
\end{equation}
Usually, the matrix $U$ is fully occupied and contains only few zeroes. Thus it is
somewhat surprising that $\op{\bi{s}}_2\cdot\op{\bi{s}}_3$ will always be tridiagonal.
On the other hand it is a straight forward task to directly
calculate $\op{\bi{s}}_2\cdot\op{\bi{s}}_3$ in the eigenbasis of $\op{\bi{s}}_1\cdot\op{\bi{s}}_2$
if one adopts the theory of irreducible tensor operators (ITO), see, e.~g.~\cite{Silver} or,
for a recent account, \cite{SS10}. In this theory the tridiagonal form of $\op{\bi{s}}_2\cdot\op{\bi{s}}_3$
is a direct consequence of the Wigner-Eckhard theorem, since $\op{\bi{s}}_2$ is an ITO of rank one.
Moreover, it follows by the same theorem that, for $S=s$, $(\op{\bi{s}}_2\cdot\op{\bi{s}}_3)_{00}=0$
which we will derive later by different arguments.
We will not dwell upon the details of the ITO calculations and simply present the result.
In order to avoid the case distinction (\ref{D9a}),(\ref{D9b}) in the formula for the
secondary diagonals of $\op{H}_2(y)$
we will express $s$ and $S$ through $g_1$ and $g_2$, the bounds of $g$, see (\ref{D10}):
\begin{eqnarray}\label{H3a}
s&=&\frac{1}{2}(g_1+g_2),\;S=\frac{1}{2}(g_2-g_1),\;\quad\mbox{for}\quad 0\le S \le s,\\
\label{H3b}
s&=&\frac{g_2}{2},\;S=\frac{1}{2}(2g_1+g_2),\;\quad\mbox{for}\quad s\le S \le 3s\;.
\end{eqnarray}
Then
\begin{eqnarray}\label{H4a}
&&\left[\op{H}_2(y) \right]_{g,g}=\frac{1}{2}\left(g(g+1)-2s(s+1) \right)+\frac{y}{4}\left(-g (1 + g) - s (1 + s) + S (S + 1)
\right),\\
\label{H4b}
&&\left[\op{H}_2(y) \right]_{g,g+1}=\left[\op{H}_2(y) \right]_{g+1,g}\\
\label{H4c}
&&=
y\sqrt{\frac{(1+g-g_1)(1+g+g_1)(g-g_2)(g-g_1-g_2)(2+g+g_2)(2+g+g_1+g_2)}{16(1+2g)(3+2g)}}
\;.
\end{eqnarray}
The other matrix elements of $\op{H}_2(y)$ vanish. For example, if $s=S=3$ then
$\op{H}_2(y)$ has the matrix form
\begin{equation}\label{H5}
\left(
\begin{array}{ccccccc}
 -12 & 4 \sqrt{3} y & 0 & 0 & 0 & 0 & 0 \\
 4 \sqrt{3} y & -11-\frac{y}{2} & \frac{3 \sqrt{15} y}{2} & 0 & 0 & 0 & 0 \\
 0 & \frac{3 \sqrt{15} y}{2} & -9-\frac{3 y}{2} & 6 \sqrt{\frac{5}{7}} y & 0 & 0 & 0 \\
 0 & 0 & 6 \sqrt{\frac{5}{7}} y & -6-3 y & \frac{11 y}{\sqrt{7}} & 0 & 0 \\
 0 & 0 & 0 & \frac{11 y}{\sqrt{7}} & -2-5 y & \frac{10 y}{\sqrt{11}} & 0 \\
 0 & 0 & 0 & 0 & \frac{10 y}{\sqrt{11}} & 3-\frac{15 y}{2} & \frac{3}{2} \sqrt{\frac{13}{11}} y \\
 0 & 0 & 0 & 0 & 0 & \frac{3}{2} \sqrt{\frac{13}{11}} y & 9-\frac{21 y}{2}
\end{array}
\right).
\end{equation}
We note in passing that a negative sign of the square root in (\ref{H4c})
gives the matrix representation of $\op{\tilde{H}}_2(y)\equiv
\op{\bi{s}}_1\cdot\op{\bi{s}}_2+y\,\op{\bi{s}}_1\cdot\op{\bi{s}}_3$. This will be later used in section \ref{sec:T}
to derive (\ref{T2a}).\\

The diagonalization of tridiagonal matrices is considerably simpler than for general matrices
and has been extensively studied also from a numerical point of view, see, e.~g.~\cite{BW}.
For the related problem of the inversion of tridiagonal matrices see \cite{HC}.
We have collected some results which will be of relevance for our purpose in the Appendix A.
For example, in our problem the entries (\ref{H4c}) of the secondary diagonals do not vanish. In this case
the eigenvalues of the tridiagonal matrix will never be degenerate. Of course, there is the usual degeneracy
due to rotational invariance and some
accidental degeneracy between the energy eigenvalues of $\op{H}$ of different invariant subspaces
$\mathcal{H}(S)$, not to mention the special case of $J_1=J_2=J_3$.\\
Another interesting property is the following : If we perturb a diagonal matrix by a tridiagonal perturbation
the resulting perturbation series will be given by relatively simple formulas,
at least for the first few terms, see (\ref{A5a})-(\ref{A5c}).
These formulas seem to be of little use at first sight since, at least for small $s$,
the eigenvalues can be numerically calculated very fast,
but we will present different applications in the sections \ref{sec:M} and \ref{sec:T}.

%%%%%%%%%%%%%%%%%%%%%%%%%%%%%%%%%%%%%%%%%%%%%%%%%%%%%%%%%%%%%%%%%%%%%%%%%%%%%%%%%%%%%%%%%%%%%%%%%%%%%%%%%%%%%%%%%%%%%%%%%%%%%%%%%%%%%%%%%%%%%%%%

\section{Spectral symmetries\label{sec:S}}

%%%%%%%%%%%%%%%%%%%%%%%%%%%%%%%%%%%%%%%%%%%%%%%%%%%%%%%%%%%%%%%%%%%%%%%%%%%%%%%%%%%%%%%%%%%%%%%%%%%%%%%%%%%%%%%%%%%%%%%%%%%%%%%%%%%%%%%%%%%%%%%%
For given quantum numbers $s$ and $S$  let $p(x,y)$ denote the characteristic polynomial
of $\op{H}_2(y)$ restricted to $\mathcal{H}(S)$, that is,
\begin{equation}\label{S1}
p(x,y)\equiv\det \left(\op{H}_2(y)-x\;\; \op{\Eins}_{\;\mathcal{H}(S)}\right)
\;.
\end{equation}
It is a polynomial in the variables $x$ and $y$ of total degree $N\equiv\dim(\mathcal{H}(S))$.
For example, if $s=S=3/2$ we have $N=4$ and
\begin{eqnarray}\nonumber
p(x,y)&=&-\frac{135}{256}(y-1)^2(33+94y+33y^2)\\
\nonumber
&&-\frac{9}{16}(y+1)(47-110y+47y^2)x\\
\nonumber
&&-\frac{9}{8}(1-26y+y^2)x^2\\
\nonumber
&&+5(1+y)x^3\\
\label{S2}
&&+x^4
\;.
\end{eqnarray}
Obviously, the above coefficients of $y^\nu\,x^\mu$ are symmetric under the reflection
$\nu\mapsto N-\mu-\nu$ or
\begin{equation}\label{S3}
y^N\,p\left(\frac{x}{y},\frac{1}{y}\right)\,=\,p(x,y) \mbox{ for } y\neq 0\;.
\end{equation}
This can be proven generally as follows.\\
The two Hamiltonians $\op{H}_2(y)=\op{\bi{s}}_1\cdot\op{\bi{s}}_2\,+y\,\op{\bi{s}}_2\cdot\op{\bi{s}}_3$
and $y\,\op{H}_2\left(\frac{1}{y}\right))=y\left(\op{\bi{s}}_1\cdot\op{\bi{s}}_2\,+\frac{1}{y}\,\op{\bi{s}}_2\cdot\op{\bi{s}}_3 \right)$
are unitarily equivalent under the permutation $(13)$ and hence have the same eigenvalues, say, $e_\nu(y),\,\nu=1,\ldots,N$.
This means that, upon an appropriate ordering of eigenvalues,
\begin{equation}\label{S4}
e_\nu(y)=y\,e_\nu\left(\frac{1}{y}\right)\;,
\end{equation}
from which (\ref{S3}) follows by
\begin{eqnarray}\label{S5a}
p(x,y)&=& \prod_{\nu=1}^N (e_\nu(y)-x)\\
\label{S5b}
&=&\prod_{\nu=1}^N \left(y\,e_\nu\left(\frac{1}{y}\right)-x\right)\\
\label{S5c}
&=&y^N\,\prod_{\nu=1}^N (e_\nu\left(\frac{1}{y}\right)-\frac{x}{y})\\
\label{S5d}
&=&y^N\,p\left(\frac{x}{y},\frac{1}{y}\right)
\;.
\end{eqnarray}
The coefficients of $y^0\,x^\mu$  are determined by the well-known
eigenvalues of $\op{H}_2(0)$
\begin{equation}\label{S6}
e_g(0)=\frac{1}{2}\left( g(g+1)-2s(s+1)\right),\;\;g=g_1,\ldots,g_2\;.
\end{equation}
For example, the factors in front of the lines of (\ref{S2}) result from
\begin{eqnarray}\nonumber
&&\left(-\frac{15}{4}-x \right)\left(-\frac{11}{4}-x \right)\left(-\frac{3}{4}-x \right)\left(\frac{9}{4}-x \right)\\
\label{S7}
&=&-\frac{135\cdot 33}{256}-\frac{9\cdot 47}{16}\,x-\frac{9}{8}\,x^2+5\, x^3+x^4\;.
\end{eqnarray}

\begin{center}
\begin{figure}
\begin{center}
\includegraphics[width=12cm]{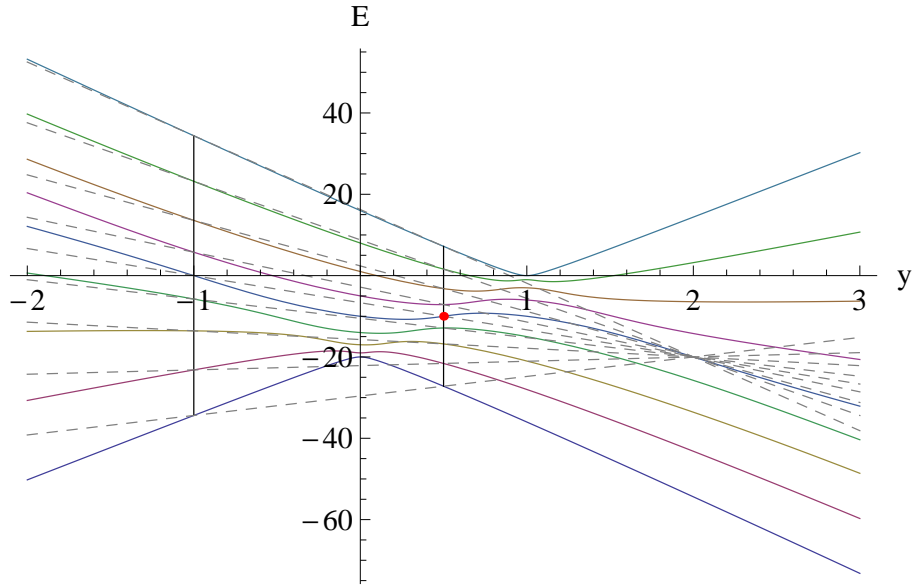}
\end{center}
\caption{\label{figG3}Spectral curves $e_\nu(y),\;\nu=0,\ldots,8$ of a $2$-chain
with coupling coefficients $J_1=1$ and $J_2=y$ and $s=S=4$.
The system of spectral curves is point-symmetric w.~r.~t.~the red point $P=(1/2,-10)$.
For pairs of $y$-values connected by the transformation (\ref{S8}) the corresponding eigenvalues
$e_\nu(y)$ are related by an affine transformation. This is illustrated here for the pair $y_1=1/2,\;y_2=1-\frac{1}{y_1}=-1$
by demonstrating that
the (dashed) lines joining the corresponding eigenvalues $e_\nu(y)$ meet at the point $(2,-20)$.
}
\end{figure}
\end{center}

There exist further symmetries of $p(x,y)$ which are not so obvious as the reflection symmetry.
They are revealed by employing arbitrary permutations of the spin triangle. These permutations
can be composed of transpositions $(i\,j)$ that
induce rational transformations of $y=\frac{J_1-J_2}{J_3-J_2}$ in the following way:
\begin{eqnarray}\label{S8}
y &&\stackrel{(13)}{\longrightarrow}\frac{1}{y} \stackrel{(12)}{\longrightarrow} 1-\frac{1}{y}=\frac{y-1}{y}
\stackrel{(23)}{\longrightarrow}\frac{y}{y-1}\stackrel{(13)}{\longrightarrow}1-\frac{y}{y-1}=\frac{1}{1-y}\stackrel{(12)}{\longrightarrow}1-y
\stackrel{(23)}{\longrightarrow}y
\;.
\end{eqnarray}
For almost all real numbers $y$ the permutation group $\mathcal{S}_3$ thus produces cycles
of six different numbers.
The exceptions are
$1\rightarrow 1\rightarrow 0\rightarrow \pm\infty\rightarrow \pm\infty\rightarrow 0\rightarrow 1$ and
$\frac{1}{2}\rightarrow 2\rightarrow -1\rightarrow -1\rightarrow 2\rightarrow \frac{1}{2}\rightarrow \frac{1}{2}$,
where only three different numbers are involved. In general, if $y$ runs through such a $6$-cycle the spectrum of the corresponding
Hamiltonians $\op{H}_2(y)$ will be related by affine transformations. (\ref{S4}) is a first example.
A second example which, together with (\ref{S4}) generates all remaining ones, is
\begin{equation}\label{S9}
e_\nu(1-y) = E_0 -e_\nu(y)
\;,
\end{equation}
where
\begin{equation}\label{S10}
E_0\equiv \frac{1}{2}\left(
S(S+1)-3s(s+1)
\right)
\;,
\end{equation}
and a suitable ordering of the eigenvalues is assumed.\\
Geometrically, (\ref{S9}) means that the system of spectral curves $y\mapsto e_\nu(y),\;\nu=1,\ldots,N$
in the $y-E$-plane is point-symmetric w.~r.~t.~the point $P\equiv (y=\frac{1}{2}, E=\frac{1}{2}E_0)$.
For odd $N$, the curve $y\mapsto e_{\frac{N+1}{2}}(y)$ passes through $P$ and is point-symmetric in itself.
Especially, $E=\frac{1}{2}E_0$ is always an eigenvalue of $\op{H}_2(\frac{1}{2})$ for odd $N$ and the
other eigenvalues of $\op{H}_2(\frac{1}{2})$ lie symmetric to the value $\frac{1}{2}E_0$ for all $N$.
These properties are illustrated in figure \ref{figG3}.

\begin{center}
\begin{figure}
\begin{center}
\includegraphics[width=12cm]{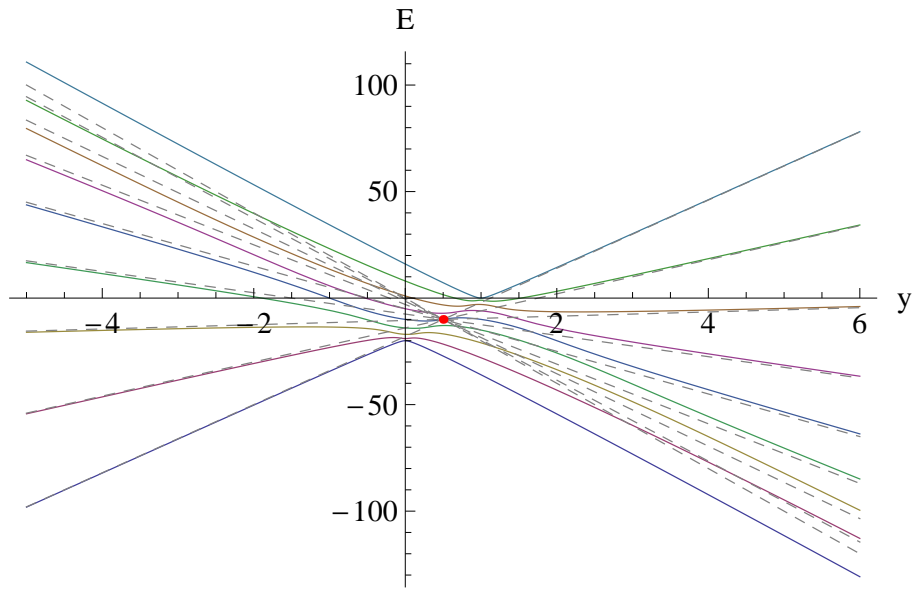}
\end{center}
\caption{\label{figG4}The same system of spectral curves as in figure \ref{figG3}, but with a larger
$y$-domain.
For large $|y|$ the spectral curves approach certain (dashed) lines which are determined by the first two terms
of the perturbation series of $e_\nu(y)$ via (\ref{S14b}). For $y\rightarrow +\infty$ the slope
of the asymptotic line equals the unperturbed eigenvalue $e_\nu(0)$; for $y\rightarrow -\infty$
the slope is $e_{N-\nu}(0)$. For $y>0$ the convergence of the spectral curves
to their asymptotic lines becomes slower for the lower curves due to larger values of $|x_2^{(\nu)}|$.
The lowest spectral curve with a large (negative) curvature at $y=0$ shows the slowest convergence.
For $y<0$ analogous remarks applies to the upper curves.\\
The asymptotic lines meet at the red point $P=(1/2,-10)$.
}
\end{figure}
\end{center}
A related consequence of (\ref{S9}) is that the shifted characteristic polynomial
\begin{eqnarray}\label{S11a}
q(x,y)&\equiv& p\left(x+\frac{E_0}{2},y+\frac{1}{2} \right)
\equiv \sum_{\nu=0}^N
q_\nu(y)\,x^\nu
\end{eqnarray}
will contain only even or odd sub-polynomials $q_\nu(y)$. For example, the shifted polynomial
pertaining to (\ref{S2}) reads:
\begin{eqnarray}\nonumber
q(x,y)&=&\frac{1}{256}(945-5130\, y^2 -4455\, y^4)
-\frac{x}{16}(27\, y+423\, y^3)\\
\label{S12}
&-&\frac{x^2}{8}(63+9\,y^2)+5\,x^3\,y + x^4
\;.
\end{eqnarray}
Unfortunately, these and the above-mentioned symmetries (\ref{S4}) are not
sufficient to determine the coefficients of $p(x,y)$ completely.\\

We will explore some further consequences of (\ref{S4}) and (\ref{S9}). For $|y|\rightarrow\infty$
the spectral curves $y\mapsto e_\nu(y)$ will approach certain lines. This can be easily understood by inserting
the perturbation series
\begin{equation}\label{S13}
e_\nu(z)=x_0^{(\nu)}+x_1^{(\nu)}\,z+x_2^{(\nu)}\,z^2+\ldots
\end{equation}
into the r.~h.~s.~of (\ref{S4}) and setting $z=\frac{1}{y}$
\begin{eqnarray}\label{S14a}
e_\nu(y)&=& y\left(  x_0^{(\nu)}+x_1^{(\nu)}\,\frac{1}{y}+x_2^{(\nu)}\,\frac{1}{y^2}+\ldots \right)\\
\label{S14b}
&=&  x_0^{(\nu)}\, y +x_1^{(\nu)}+x_2^{(\nu)}\,\frac{1}{y}+\ldots
\;,\quad \mbox{ for }y\rightarrow +\infty\;.
\end{eqnarray}
For $y\rightarrow -\infty$ one obtains
\begin{equation}\label{S14c}
e_\nu(y)= x_0^{(N-\nu)}\, y +x_1^{(N-\nu)}+x_2^{(N-\nu)}\,\frac{1}{y}+\ldots
\;.
\end{equation}
Since the spectral curves do not intersect
(see Appendix A), it follows that for $y\rightarrow +\infty$ the spectral curve
$e_\nu(y)$ approaches the line with slope $x_0^{(\nu)}$ and for $y\rightarrow +\infty$ it
approaches the line with slope $x_0^{(N-\nu)}$. This effect is illustrated in figure \ref{figG4}.\\
Anticipating the result (\ref{S16a}) it can be shown that the $N$ asymptotic lines
meet at the point $P= (y=\frac{1}{2}, E=\frac{1}{2}E_0)$:
\begin{equation}\label{S15a}
x_0^{(\nu)}\, \frac{1}{2} +x_1^{(\nu)}
\stackrel{(\ref{S16a})}{=}
x_0^{(\nu)}\, \frac{1}{2}+\frac{1}{2}\left( E_0 -x_0^{(\nu)}\right)
=\frac{1}{2}\, E_0
\;.
\end{equation}

By the sequence of rational maps (\ref{S8}) the closed interval $[0,\frac{1}{2}]$ is mapped as follows:
\begin{equation}\label{S15}
[0,\frac{1}{2}]\rightarrow [2,+\infty]\rightarrow [-\infty,-1] \rightarrow[-1,0]
\rightarrow [1,2] \rightarrow [\frac{1}{2},1] \rightarrow [0,\frac{1}{2}]
\;.
\end{equation}
This set of six closed intervals covers the whole real axis, hence the values of $e_\nu(y)$
for $y\in [0,\frac{1}{2}]$ completely determine the full spectral curve $e_\nu(y),\,y\in\mathbb{R}$.
We have already exploited this principle for the discussion of the asymptotic behavior of $e_\nu(y)$ for $|y|\rightarrow\infty$.\\
Conversely, if we choose some analytic function, restrict it to $y\in [0,\frac{1}{2}]$ and extend it to the full
real line by means of the functional equations (\ref{S4}) and (\ref{S9}), the resulting function will not automatically
be analytic at the borders of the intervals, e.~g.~at $y=0$.
This makes it plausible that (\ref{S4}) and (\ref{S9}) impose conditions on the perturbation
series (\ref{S13}). Let us consider some small $y$. After the first three steps of the sequence (\ref{S8})
$y$ is mapped onto $\frac{-y}{1-y}$ which is again small. Simultaneously, $e_\nu(y)$ is mapped
onto
\begin{eqnarray}\label{S15a}
E_0-\frac{1}{1-y}\left( E_0-e_\nu(y)\right)
&=&-E_0 \frac{y}{1-y}+\frac{1}{1-y}\sum_{i=0}^{\infty}x_i^{(\nu)} y^i\\
\label{S15b}
&=&e_\nu\left( \frac{-y}{1-y}\right)
=\sum_{i=0}^{\infty}x_i^{(\nu)}\left( \frac{-y}{1-y}\right)^i
\;.
\end{eqnarray}
Equating the coefficients of the power series (\ref{S15a}) and (\ref{S15b}) leads to the following
system of equations:
\begin{equation}\label{S16}
0=-E_0+\sum_{n=0}^m \left( 1+(-1)^{n+1} {m-1 \choose n-1}\right) x_n^{(\nu)},\quad m=1,2,\ldots
\end{equation}
Consequently,
for all $\nu=1,\ldots,N$
the perturbation coefficients $x_n^{(\nu)}$, $n$ odd, can be expressed in terms of the perturbation coefficients $x_i^{(\nu)}$,
$i$ even and $i<n$. The first few of the corresponding equations read:
\begin{eqnarray}\label{S16a}
x_1^{(\nu)}&=& \frac{1}{2}\left( E_0 -x_0^{(\nu)}\right)\\
\label{S16b}
x_3^{(\nu)}&=& \frac{1}{2}x_2^{(\nu)}\\
\label{S16c}
x_5^{(\nu)}&=& \frac{1}{4}\left( -x_2^{(\nu)}+6 x_4^{(\nu)}\right)\\
\label{S16d}
x_7^{(\nu)}&=& \frac{1}{2}\left( x_2^{(\nu)} -5x_4^{(\nu)}+5x_6^{(\nu)}\right)\\
\label{S16e}
x_9^{(\nu)}&=& -\frac{17}{8}x_2^{(\nu)}+\frac{21}{2}x_4^{(\nu)}-\frac{35}{4}x_6^{(\nu)}+\frac{7}{2}x_8^{(\nu)}
\;.
\end{eqnarray}
Sometimes we will switch from the index notation of eigenvalues using $\nu=1,\ldots,N$ to the notation using $g=g_1,\ldots,g_2$
introduced before. In this notation we have $x_0^{(g)}=\frac{1}{2}(g(g+1)-2s(s+1))$. In the special case of
$S=s$ we conclude $E_0=\frac{1}{2}(S(S+1)-3s(s+1))=-s(s+1)$ and, by (\ref{S16a}),  $x_1^{(g)}=-\frac{1}{4}g(g+1)$.
Especially it follows that $x_1^{(0)}=0$, which was previously, in section \ref{sec:H},
mentioned as a consequence of the Wigner-Eckhard theorem. This fact will simplify some calculations in section \ref{sec:M}.
There are two other special cases that should be mentioned:
One case is $S=0$ with $N=1$  for integer $s$. The corresponding eigenstate does not depend on $y$, see \cite{SR10},
hence the spectral curve coincides with its asymptotic line.\\
The other case is $s=1/2$ and the vector ${\mathbf J}=(J_1,J_2,J_3)$ running through a circle in the plane $J_1+J_2+J_3=3j=\mbox{const}$
with center $(j,j,j)$. In this case the eigenvalues of $\op{H}$ will be constant, i.~e.~, we have the rare case of isospectrality
for spin systems, see \cite{SL2001}.\\

As a further application of the spectral symmetry we calculate the first five moments of $\op{H}_2(y)$.
Let
\begin{equation}\label{S17}
M_n(y)\equiv \sum_{\nu=1}^N \left(e_\nu(y)  \right)^n,\;n\in\mathbb{N}\;.
\end{equation}
Since the moments can be expressed through the coefficients of the characteristic polynomial, see, e.~g.~, \cite{Aldrovandi} (14.35),
the $M_n(y)$ will be polynomials in the variable $y$ of maximal degree $n$ and their coefficients will exhibit the same
reflection symmetry as the coefficients of $p(x,y)$, see (\ref{S4}). For example,
\begin{equation}\label{S18}
M_1(y)=M_{10}(1+y)
\;.
\end{equation}
The coefficient $M_{10}$ can be calculated directly or by using
\begin{eqnarray}\label{S19a}
M_1(1-y)&=& \sum_{\nu=1}^N (E_0-e_\nu(y))\\
\label{S19b}
&=& N\,E_0-M_1(y)\\
\label{S19c}
&=& N\,E_0-M_{10}(1+y)\\
\label{S19d}
&=& M_{10}(1+1-y)
\;,
\end{eqnarray}
hence
\begin{equation}\label{S20}
M_{10}= \frac{1}{3}\,N\,E_0
\;.
\end{equation}

Similarly,
\begin{eqnarray}\label{S21a}
M_2(y)&=&M_{20}(1+y^2)+M_{21}y\;,\\
\label{S21b}
M_2(1-y)&=& (2 M_{20}+M_{21})(1-y)+M_{20}y^2\\
\label{S21c}
&=& \sum_{\nu=1}^N (E_0-e_\nu(y))^2\\
\label{S21d}
&=& N\,E_0^2-2 E_0\,M_1(y)+M_2(y)\\
\label{S21e}
&=& N E_0^2-2 E_0 M_{10}(1+y)+M_{20}(1+y^2)+M_{21}y
\;,
\end{eqnarray}
hence
\begin{equation}\label{S22}
M_2(y)= M_{20}(1+y^2)+(E_0 M_{10}-M_{20})y
\;.
\end{equation}
Here and in the following equations we assume the  $M_{n0}$  as given
since they can be expressed through the well-known eigenvalues of $\op{H}_2(0)$.
By the same method we obtain
\begin{equation}\label{S23}
M_3(y)= M_{30}(1+y^3)+\frac{3}{2}(E_0 M_{20}-M_{30})y(1+y)
\;,
\end{equation}
\begin{eqnarray}\nonumber
M_4(y)&=& M_{40}(1+y^4)+2(E_0 M_{30}-M_{40})y(1+y^2)\\
\label{S24}
&+&\left(
-E_0^3 M_{10}+6E_0^2 M_{20}-8E_0M_{30}+3 M_{40}
\right)y^2
\;,
\end{eqnarray}
and
\begin{eqnarray}\nonumber
M_5(y)&=& M_{50}(1+y^5)+\frac{5}{2}(E_0 M_{40}-M_{50})y^2(1+y^2)\\
\label{S25}
&+&\left(
-E_0^4 M_{10}+5E_0^3 M_{20}-5E_0^2 M_{30}+ M_{50}
\right)y^3
\;.
\end{eqnarray}
For $n\ge 6$ this method fails to give $M_{n2}$ in terms of $M_{k0},\;k\le n$
and simple formulas like those above seem to be no longer available. Note, however, that the
first five moments have been derived solely by utilizing $y=0$ results and spectral symmetry.
The knowledge about the explicit matrix representation of $\op{H}_2(y)$ has not yet been used.\\
The first
five moments of $\op{H}_2(y)$ depend on $S$ only implicitly through $E_0$
and $M_{n0}$. If these arguments are explicitly evaluated it is possible to
calculate the corresponding total moments of the $2$-chain by summation over $S$ and to compare the
results with those in the literature, see \cite{RBW} and \cite{SLR}. This has been done
successfully but will not be detailed here. Rather, in the next two sections, we will use an alternate
method to calculate moments which leads to results not already reported in the literature.

%%%%%%%%%%%%%%%%%%%%%%%%%%%%%%%%%%%%%%%%%%%%%%%%%%%%%%%%%%%%%%%%%%%%%%%%%%%%%%%%%%%%%%%%%%%%%%%%%%%%%%%%%%%%%%%%%%%%%%%%%%%%%%%%%%%%%%%%%%%%%%%%

\section{Moments of the two-chain\label{sec:M}}

%%%%%%%%%%%%%%%%%%%%%%%%%%%%%%%%%%%%%%%%%%%%%%%%%%%%%%%%%%%%%%%%%%%%%%%%%%%%%%%%%%%%%%%%%%%%%%%%%%%%%%%%%%%%%%%%%%%%%%%%%%%%%%%%%%%%%%%%%%%%%%%%
The total $k$-th moment of $\op{H}_2(y)$ will be defined as
\begin{equation}\label{M1}
t_k(y)\equiv \frac{\mbox{Tr }\op{H}_2(y)^k}{(2s+1)^3}
\;.
\end{equation}
Obviously, it is a polynomial in the variable $y$ with the same reflection symmetry
as the local moments $M_k(y)$ defined in (\ref{S17}). The knowledge of the first
terms in the perturbation series of the eigenvalues $e_{g,S}(y)$ of $\op{H}_2(y)$ will
give the corresponding coefficients of $t_k(y)$ via summation over the quantum numbers
$S$ and $g\equiv S_{12}$.
Moreover, we will see that there is a close connection between these coefficients
and the terms of the so-called graph expansion of $t_k(y)$. In the case of the two-chain
$\op{H}_2(y)$ this connection is even a $1:1$-correspondence.\\
Let
\begin{equation}\label{M2}
e_{g,S}(y)=x_0^{(g)}+x_1^{(g)}\,y+x_2^{(g)}\,y^2+x_3^{(g)}\,y^3+\ldots
\end{equation}
denote the perturbation series of the eigenvalues $e_{g,S}(y)$.
Upon multinomial expansion we obtain
\begin{eqnarray}\nonumber
e_{g,S}(y)^k&=&x_0^{(g)k}+k\,x_0^{(g)(k-1)}\,x_1^{(g)}\,y\\
\nonumber
&+&\left({k\choose 2}\,x_0^{(g)(k-2)}\,x_1^{(g)2}+k\,x_0^{(g)(k-1)}\,x_1^{(g)}\right)y^2\\
\nonumber
&+&\left({k\choose 3}x_0^{(g)(k-3)}x_1^{(g)3}+k(k-1)x_0^{(g)(k-2)}x_1^{(g)}x_2^{(g)}
+k x_0^{(g)(k-1)}x_3^{(g)}\right)y^3\\
\nonumber
\\
\label{M3a}
&+&\ldots
\end{eqnarray}
Into this series we insert the known equations (\ref{S16a}) and (\ref{S16b}):
\begin{eqnarray}\label{M4a}
x_0^{(g)}&=&\frac{1}{2}\left(g(g+1)-2s(s+1)\right)\;,\\
\label{M4b}
x_1^{(g)}&=&\frac{1}{2}(E_0-x_0^{(g)})=\frac{1}{2}(S(S+1)-g(g+1)-s(s+1))\;,\\
\label{M4c}
x_3^{(g)}&=&\frac{1}{2}x_2^{(g)}
\;.
\end{eqnarray}
For $x_2^{(g)}$ we will use the general result (\ref{A5a}) for tridiagonal matrices and
the explicit form (\ref{H4c})  of the secondary diagonal entries of $\op{H}_2(y)$.
This yields, after some calculations, for $g>0$:
\begin{eqnarray}\nonumber
x_2^{(g)}&=&\frac{1}{16 (2 g+1)}\\
\nonumber
&&
\left(
\frac{(g-2 s-1) (g+2 s+1) (g-s-S-1) (g+s-S) (g-s+S) (g+s+S+1)}{g (2 g-1)}
\right.\\
\nonumber
&-&\left.\frac{(g-2 s) (g+2 s+2) (g-s-S) (g+s-S+1)(g-s+S+1) (g+s+S+2)}{(g+1) (2 g+3)}
\right)\\
&&\label{M5a}\\
\label{M5b}
&=&-\frac{3}{32} x_0^{(g)} + \frac{E_0}{16} + \frac{3}{256} (1 + 8 s(s+1)) + \frac{A_1}{g (g + 1)} -\frac{4 A_2}{4g (g + 1)-3}
\;,
\end{eqnarray}
where
\begin{equation}\label{M6}
A_1=\frac{1}{16} (1 + 2 s)^2 (s - S)^2 (1 + s + S)^2
\;,
\end{equation}
and
\begin{equation}\label{M7}
A_2=\frac{1}{1024} (1 + 4 s) (3 + 4 s) (-1 + 2 s - 2 S) (1 + 2 s - 2 S) (1 + 2 s + 2 S) (3 + 2 s + 2 S)
\;.
\end{equation}
For $g=0$ the first term in the brackets in (\ref{M5a}) or, equivalently, the term $\frac{A_1}{g (g + 1)}$ in (\ref{M5b})
has to be set to $0$.
These results have to be inserted into
\begin{equation}\label{M8}
t_k(y)=\sum_{g=0}^{2s}\sum_{S=|s-g|}^{s+g}(2S+1)e_{g,S}(y)^k
\;.
\end{equation}
The double sum in (\ref{M8}) can be evaluated up to third order in $y$ by adopting the following strategy:
First, expand all terms depending on $S(S+1)$ into polynomials in $E_0$ and express the resulting sums of the form
\begin{equation}\label{M9}
\sum_{S=|s-g|}^{s+g}(2S+1)E_0^m
\;,
\end{equation}
up to a factor $2g+1$, as polynomials in $x_0^{(g)}$. This makes is possible to write the
final sums $\sum_{g=0}^{2s}\ldots x_0^{(g)\lambda}$ in terms of the known moments $t_\lambda (0)$ (dimer polynomials).\\

Before writing down the final results of the above-sketched calculations we will
shortly recapitulate the general theory of moments of Heisenberg Hamiltonians $\op{H}$, following
\cite{RBW} and \cite{SLR},
that is also the starting point of the high temperature expansion of the corresponding
thermodynamic quantities. The $k$-th moment of $\op{H}$ has the following ``graph expansion"
\begin{equation}\label{M10}
t_k=\frac{\mbox{Tr }\op{H}^k}{(2s+1)^N}=\sum_{\mathcal{G}}\overline{\mathcal{G}}\,p({\mathcal{G}})
\;.
\end{equation}
The latter summation extends over a finite set of (multi-)graphs with $k$ edges (bonds).  $\overline{\mathcal{G}}$
denotes a certain polynomial of the coupling constants of $\op{H}$
that is obtained via summing over all embeddings of ${\mathcal{G}}$ into the spin system, see \cite{SLR}.
$p({\mathcal{G}})$ denotes certain other
``universal" polynomials of maximal degree $k$ in the variable $r\equiv s(s+1)$ that are independent of the
coupling constants of the spin system under consideration.\\
In our case of the two-chain $\op{H}_2(y)$ the graphs needed in the graph expansion of $t_k(y)$ are
either of the form $\;\dii$ (two vertices and $k$ edges) or $\;\tii\;$ (three vertices and a total number of $k=k_1+k_2$ edges).
The cases ${\mathcal{G}}=\;\di$ or ${\mathcal{G}}=\;\ti\;$ can be disregarded since then $p({\mathcal{G}})=0$, see \cite{RBW}.
Moreover, we have
$\overline{\mathcal{G}}=1+y^k$ in the first case of ${\mathcal{G}}=\;\dii\;$ and
$\overline{\mathcal{G}}=y^{k_1}+y^{k_2}$ in the second case of ${\mathcal{G}}=\;\tii\;$.
Hence the coefficients of the $y$-expansion of $t_k(y)$ are exactly the universal polynomials $p({\mathcal{G}})$
of the relevant graphs and will be written as $p(k)$ in the first case and $p(k_1,k-k_1),\;k_1\le k-k_1,$ in the second one.
The dimer polynomials $p(k)$ are the well-known moments of the one-chain and will be given in explicit form in the Appendix B.
The polynomials $p(k_1,k-k_1),\;k_1=2,3,4,5$  can be expressed in terms of dimer polynomials $p(k')$ as the result of the above-sketched
calculations. The results are
\begin{equation}\label{M11}
p(2,k-2)=\frac{k}{12}\left[
(-r)^k +(k-1)\,r^2\, p(k-2)+k\,r \,p(k-1)
\right]
\;,
\end{equation}
and
\begin{eqnarray}\nonumber
p(3,k-3)&=&\frac{1}{144}\left[
-3k(2r+k-1)(-r)^{k-1} +6k^2 r p(k-1)\right.\\
\label{M12}
&-&\left.2r(6r+k+1){k \choose 2}p(k-2)
-6 r^2 {k \choose 3}p(k-3)
\right]
\;.
\end{eqnarray}
If $k_1=4,5$ the expressions for $p(k_1,k-k_1)$ are more complex and will involve sums over ${\mathcal O}(k)$ dimer polynomials.
It will be economic to write the results in terms of the modified dimer functions
\begin{equation}\label{M13}
P(\ell)\equiv p(\ell)-\frac{(-r)^\ell}{4r+1},\quad \ell=0,1,2,\ldots
\;,
\end{equation}
although the expressions for $p(k_1,k-k_1)$ will then no longer be manifestly polynomial.
The final results are:
\begin{eqnarray}\nonumber
&&p(4,k-4)=\\ \nonumber
&&\frac{k r \left(-32 r (4 r-3) (80 r+21) \left(\frac{3}{8}-r\right)^{k-1}-128 (-r)^k (r (-30 k+80 r-9)-72)\right)}
{69120 (4r+1)}\\ \nonumber
&&+ \frac{5 k^2 r}{192}\;P(k-1)+ \frac{(k-1) k r \left(16 k^2 (3 r-1)-16 k (3 r+4)-3 (284 r+57)\right)}{23040}\;P(k-2)\\
\nonumber
&&+\frac{k \left(k^2-3 k+2\right) r \left(k \left(12 r^2-5 r+2\right)-36 r^2+39 r+6\right)}{2880}\;P(k-3)\\
\nonumber
&&+ \frac{(k-3) (k-2) (k-1) k r^2 \left(6 r^2-3 r+2\right)}{2880}\;P(k-4)\\
\nonumber
&&+\frac{k}{960} \sum _{i=0}^{k-3} (i+1)  \left(\frac{1}{64} r (4 r-3) (16 r+3)^2
\left(\frac{3}{8}-r\right)^{-i+k-3}+(r-2) (4 r+1)^2 (-r)^{-i+k-2}\right)P(i), \\
&&\label{M14}
\end{eqnarray}
and
\begin{eqnarray}\nonumber
&&p(5,k-5)=\\ \nonumber
&&4 k r^2 (4 r-3) \left(\frac{3}{8}-r\right)^{k-2}\;\frac{(80 r+21)  (4 k+8 r-7)}{46080 (4 r+1)}\\
\nonumber
&&-64 k (-r)^k\;\frac{
20 k r^2-45 k r-15 k+160 r^3-98 r^2-39 r+15}{46080 (4 r+1)}\\ \nonumber
&&+ \frac{7}{384} k^2 r P(k-1)+(k-1) k r \;\frac{ k^2 (96 r-32)+k (-96 r-68)-3 (532 r+111)}{46080}P(k-2)\\
\nonumber
&&-(k-2) (k-1) k\; \frac{ 2 k (8 r-3) r (k (2 r-1)-1)-108 r^3+372 r^2-141 r-45}{11520 (8 r-3)}P(k-3)\\
\nonumber
&&-(k-3) (k-2) (k-1) k r \;\frac{8 k r^2-5 k r+k+24 r^3-52 r^2+34 r+6}{11520}P(k-4)\\
\nonumber
&&-(k-4) (k-3) (k-2) (k-1) k r^2\;\frac{ 4 r^2-3 r+1 }{11520}P(k-5)\\
\nonumber
&&+\frac{k}{3840} \sum _{i=0}^{k-3} (i+1) \Bigg(-\frac{3}{256} r (4 r-3) (16 r+3)^2 (4 k+8 r-7)
   \Big(\frac{3}{8}-r\Big)^{-i+k-4}\\
\nonumber
&&\quad   -(4 r+1)^2 (5 k (r-1)+r (6 r-17)+5) (-r)^{-i+k-3}\Bigg) P(i). \\
&&\label{M15}
\end{eqnarray}

The formulas (\ref{M11}), (\ref{M12}), (\ref{M14}) and (\ref{M15}) have been checked with the known results for $k=2,\ldots,8$ in \cite{RBW}
and unpublished results for $k=9,10$ that have been obtained by different methods.

%%%%%%%%%%%%%%%%%%%%%%%%%%%%%%%%%%%%%%%%%%%%%%%%%%%%%%%%%%%%%%%%%%%%%%%%%%%%%%%%%%%%%%%%%%%%%%%%%%%%%%%%%%%%%%%%%%%%%%%%%%%%%%%%%%%%%%%%%%%%%%%%

\section{Moments of the general triangle\label{sec:T}}

%%%%%%%%%%%%%%%%%%%%%%%%%%%%%%%%%%%%%%%%%%%%%%%%%%%%%%%%%%%%%%%%%%%%%%%%%%%%%%%%%%%%%%%%%%%%%%%%%%%%%%%%%%%%%%%%%%%%%%%%%%%%%%%%%%%%%%%%%%%%%%%%
The graph expansion of $t_k=\frac{1}{(2s+1)^3}\mbox{Tr }\op{H}^k$ for the general triangle Hamiltonian (\ref{D1})
involves additional triangular graphs with $k_1,\,k_2,\,k_2$ bonds such that $k_1+k_2+k_3=k$.
We will write the corresponding universal polynomials as $p(k_1,k_2,k_3),\, k_1\le k_2 \le k_3$.
Following the ideas of the last section, a possible strategy to determine these polynomials would
consist of expanding $t_k(z)$ into an even power series w.~r.~t.~$z$, if the Hamiltonian (\ref{D1})
is written in the form
\begin{equation}\label{T1}
\op{H}(z)=J_3\,\op{\bi{s}}_1\cdot \op{\bi{s}}_2\
+
\,J\left(\op{\bi{s}}_2\cdot\op{\bi{s}}_3+\op{\bi{s}}_1\cdot\op{\bi{s}}_3\right)
+
\,(J_3-J)\,z\left(\op{\bi{s}}_2\cdot\op{\bi{s}}_3-\op{\bi{s}}_1\cdot\op{\bi{s}}_3\right)
\;.
\end{equation}
In this paper we will take only the first step of this expansion and consider the integrable case
$z=0$. Even this case is sufficient to obtain explicit expressions for some classes of universal
polynomials.\\
Consider the eigenvalues $e_{g,S}$ of $\op{H}(0)$ that, according to (\ref{S16a})
and the remark following (\ref{H5}),  can be written as
\begin{eqnarray}\label{T2a}
e_{g,S}&=&J_3\,x_0^{(g)}\,+\,2 J\, x_1^{(g)}\\
\label{T2b}
&=&J_3\,x_0^{(g)}\,+ J\, (E_0-x_0^{(g)})
\;.
\end{eqnarray}
Hence
\begin{eqnarray}\label{T3a}
e_{g,S}^k&=&\sum_{\ell=0}^k\, \binom{k}{\ell}\,J_3^\ell\, x_0^{(g)\ell}\, J^{k-\ell} \,(E_0-x_0^{(g)})^{k-\ell}\\
\label{T3b}
&=&\sum_{\ell=0}^k \,\sum_{\mu=0}^{k-\ell}\, \binom{k}{\ell}\,\binom{k-\ell}{\mu}\, J_3^\ell \,J^{k-\ell}
E_0^\mu(-1)^{k-\ell-\mu}\,x_0^{(g)(k-\mu)}
\;.
\end{eqnarray}
The moments $t_k(0)$ are obtained by summing (\ref{T3b}) over $S$ and $g$. This leads to sums of the form
$\sum_{S=|s-g|}^{s+g} (2S+1)E_0^\mu$. It can be shown that these are of the form $(2s+1)(2g+1)$ times
a polynomial in the variables $r=s(s+1)$ and $g(g+1)$. Hence we obtain the following expansion
\begin{equation}\label{T4}
\sum_{S=|s-g|}^{s+g} (2S+1)E_0^\mu = (2s+1) (2g+1) \sum_{\lambda=0}^\mu
P_{\mu\lambda}\,x_0^{(g)\lambda}
\;,
\end{equation}
which implicitly defines the polynomials $P_{\mu\lambda}(r)$. For our applications we will
only use the following ones:
\begin{eqnarray}\label{T5a}
P_{nn}&=&1,\;n=0,1,2,\ldots,\\
\label{T5b}
P_{10}&=&0,\\
\label{T5c}
P_{20}&=&\frac{2}{3}r^2,\;P_{21}=\frac{2}{3}r,\\
\label{T5d}
P_{30}&=&-\frac{1}{3}r^2,\;P_{31}=\frac{1}{3}r(6r-1),\;P_{32}=2r
\;.
\end{eqnarray}
Note that the factor $(2s+1)(2g+1)$ is just the degeneracy of the eigenvalue $x_0^{(g)}$ of
$\op{\bi{s}}_1\cdot \op{\bi{s}}_2$ in the total Hilbert space. Hence the sums
of the form $\sum_{g=0}^{2s}\,(2s+1)(2g+1)\,x_0^{(g)\lambda}$ are equal to
$(2s+1)^3 p(\lambda)$ and we obtain
\begin{equation}\label{T6}
t_k(0)=\sum_{\ell=0}^k \sum_{\mu=0}^{k-\ell}\sum_{\lambda=0}^{\mu}
\binom{k}{\ell}\binom{k-\ell}{\mu}
J_3^\ell J^{k-\ell}(-1)^{k-\ell-\mu}
P_{\mu\lambda}\,p(k+\lambda-\mu)
\;.
\end{equation}
We will only consider two terms of the expansion (\ref{T6}),
namely those with $\ell=k-2$ and $\ell=k-3$. Let $[A(x)]_x$
denote the coefficient of $x$ in the polynomial $A(x)$.
Then the first term is
\begin{eqnarray}\label{T7a}
[t_k(0)]_{J_3^{k-2}J^2}
&=&
\binom{k}{2}\sum_{\mu=0}^2 \sum_{\lambda=0}^\mu \binom{2}{\mu}(-1)^{2-\mu}P_{\mu\lambda}\,p(k+\lambda-\mu)\\
\label{T7b}
&=&p(1,1,k-2)+2p(2,k)\;.
\end{eqnarray}
After some calculations, using (\ref{M11}) and (\ref{M12}), we finally obtain
\begin{eqnarray}\nonumber
p(1,1,k-2)&=&\frac{k}{6}\left[
-(-r)^k+(k-1)r^2
p(k-2)+(k-2)p(k-1)\right],\\ \label{T8}
&& \mbox{ for }
k=3,4,\ldots\; .
\end{eqnarray}
For $\ell=k-3$ the analogous results read
\begin{eqnarray}\label{T9a}
[t_k(0)]_{J_3^{k-3}J^3}
&=&
\binom{k}{3}\sum_{\mu=0}^3 \sum_{\lambda=0}^\mu \binom{3}{\mu}(-1)^{3-\mu}P_{\mu\lambda}\,p(k+\lambda-\mu)\\
\label{T7b}
&=&2\left(p(1,2,k-3)+2p(3,k)\right)\;.
\end{eqnarray}
and
\begin{eqnarray}\nonumber
p(1,2,k-3)&=&\frac{k}{48}  \left\{
(k+2 r-1) (-r)^{k-1}\right.\\
\nonumber
&+&r \left[(k-1)\, \left((2-k)\, r \,p(k-3)\,+\, (-k+2 r+3)\,p(k-2)\right)\right.\\
\label{T10}
&-&\left.\left.2\, k \,p(k-1)\right]
\right\},\quad\mbox{ for }
k=4,5,\ldots\; .
\end{eqnarray}
The formulas (\ref{T8}) and (\ref{T10}) agree with the values of the polynomials published in \cite{RBW} for $k=4,\ldots,8$
and with unpublished results for $k=9,10\;$ obtained by different methods. For an example see Appendix B.\\

It should be mentioned that the universal polynomials (\ref{M11}), (\ref{M12}), (\ref{M14}), (\ref{M15}), (\ref{T8}), and (\ref{T10})
do not only occur in the graph expansion of the moments of the two-chain or the general triangle
but for arbitrary Heisenberg spin systems. As a rule, general formulas for these polynomials are rare.
The only ones previously known to the author are, besides the explicit expressions for the dimer polynomials $p(k)$, see Appendix B,
those which result from certain rules given in \cite{RBW}. For example, $p({\mathcal{G}})=0$ if ${\mathcal{G}}$
is a single-bond connection of two disjoint (possibly empty) graphs. Another rule in \cite{RBW}
refers to the case where ${\mathcal{G}}$ contains two vertices connected by a single-bond chain of length $n>2$.
Let ${\hat{\mathcal{G}}}$ be the graph resulting by shortening the single-bond chain by one vertex.
Then $p({\mathcal{G}})=\frac{\ell\,r}{3}p({\hat{\mathcal{G}}})$, where $\ell$ is the total number of bonds of ${\mathcal{G}}$.
(Note that the authors of \cite{RBW} use a definition of the graph polynomials $p({\mathcal{G}})$ that differs
from our definition by a multinomial factor).
In particular, the last rule implies that
the polynomial $q_n$ of a single-bond polygon with $n>2$ vertices will be $q_n(r)=\frac{n!}{3^{n-1}}r^n$.
Of course, this rule can also be applied to graphs that can ultimately be reduced to those for which
(\ref{M11}), (\ref{M12}), (\ref{M14}), (\ref{M15}), (\ref{T8}), and (\ref{T10}) apply.\\

%%%%%%%%%%%%%%%%%%%%%%%%%%%%%%%%%%%%%%%%%%%%%%%%%%%%%%%%%%%%%%%%%%%%%%%%%%%%%%%%%%%%%%%%%%%%%%%%%%%%%%%%%%%%%%%%%%%%%%%%%%%%%%%%%%%%%%%%%%%%%%%%

\section{Summary and outlook\label{sec:SO}}
%%%%%%%%%%%%%%%%%%%%%%%%%%%%%%%%%%%%%%%%%%%%%%%%%%%%%%%%%%%%%%%%%%%%%%%%%%%%%%%%%%%%%%%%%%%%%%%%%%%%%%%%%%%%%%%%%%%%%%%%%%%%%%%%%%%%%%%%%%%%%%%%
The general quantum spin triangle is not integrable in the strict sense that its eigenvalues and eigenvectors can be given
in closed form, as, e.~g.~, for the isosceles triangle. However, our results show that we are rather close to integrability.
The reduced Hamiltonian has a triangular matrix representation the entries of which are explicitly given. Hence also the
characteristic polynomial can be expressed either recursively or explicitly in terms of finite sums, see Appendix A.
Similarly, the perturbation series for the eigenvalues could, in principle, be calculated up to any finite order,
although with rapidly increasing complexity.
Therefore we have contented ourselves with third order perturbation results.\\
Comparing this situation with, e.~g.~, the Bethe ansatz for the $s=\frac{1}{2}$ chain with $N$ spins shows that the difference is not huge.
Also in the Bethe ansatz the eigenvalues have to be calculated as the numerical solutions of certain equations which become
more and more complex for large $N$. Perhaps the general triangle should also be regarded as ``quantum integrable" in some weaker sense.\\
One virtue of integrability in whatever sense is the possibility to calculate certain limits, e.~g.~, the thermodynamic
limit $N\rightarrow \infty$ in the case of the $s=\frac{1}{2}$ chain. In the present case a possible candidate is the classical
limit $s\rightarrow \infty$, see \cite{MSSL} for the equilateral triangle case. This seems to be a sensible project for future work.
Another aspect that has only be touched in the present paper is the closer investigation of physical properties of the general spin
triangle, as, for instance, ground state diagrams or magnetization curves. As already pointed out in the Introduction, these properties
could be of paradigmatic character in order to understand more complex systems.

%%%%%%%%%%%%%%%%%%%%%%%%%%%%%%%%%%%%%%%%%%%%%%%%%%%%%%%%%%%%%%%%%%%%%%%%%%%%%%%%%%%%%%%%%%%%%%%%%%%%%%%%%%%%%%%%%%%%%%%%%%%%%%%%%%%%%%%%%%%%%%%%

\section*{Acknowledgement}
%%%%%%%%%%%%%%%%%%%%%%%%%%%%%%%%%%%%%%%%%%%%%%%%%%%%%%%%%%%%%%%%%%%%%%%%%%%%%%%%%%%%%%%%%%%%%%%%%%%%%%%%%%%%%%%%%%%%%%%%%%%%%%%%%%%%%%%%%%%%%%%%
To check the equations (\ref{M11}), (\ref{M12}), (\ref{M14}), (\ref{M15}), (\ref{T8}), and (\ref{T10}) I have used unpublished results for $k=9,10$
that have been obtained in collaboration with Andre Lohmann and Johannes Richter. I thank Johannes Richter and J\"urgen Schnack for
critical reading of the first version of the manuscript and Thomas Br\"ocker for encouraging discussions.

%%%%%%%%%%%%%%%%%%%%%%%%%%%%%%%%%%%%%%%%%%%%%%%%%%%%%%%%%%%%%%%%%%%%%%%%%%%%%%%%%%%%%%%%%%%%%%%%%%%%%%%%%%%%%%%%%%%%%%%%%%%%%%%%%%%%%%%%%%%%%%%%

\section*{Appendix A: Properties of tridiagonal matrices\label{sec:A}}
%%%%%%%%%%%%%%%%%%%%%%%%%%%%%%%%%%%%%%%%%%%%%%%%%%%%%%%%%%%%%%%%%%%%%%%%%%%%%%%%%%%%%%%%%%%%%%%%%%%%%%%%%%%%%%%%%%%%%%%%%%%%%%%%%%%%%%%%%%%%%%%%

We will collect some properties of tridiagonal matrices relevant for this article without claiming originality.
For sake of simplicity we assume $T$ to be an $N\times N$-symmetric real tridiagonal matrix with positive entries in the secondary diagonals.
Let $T^{(n)}$ denote the submatrix of $T$ obtained by deleting its first $n$ rows and colums and $D_n\equiv \det T^{(n)}$.
Then $D_0=\det T$ satisfies the obvious recursion relation
\begin{equation}\label{A1}
D_0=T_{11}\,D_1-T_{12}^2\,D_2
\;.
\end{equation}
It follows that the explicit formula for $\det T$ does not contain $N!$ terms as for general $T$, but only
$F_{N+1}$ terms, where $F_{N}$ denotes the $N$-th Fibonacci number.
For example, if $N=3$ we have $\det T = T_{11}T_{22}T_{33}-T_{11}T_{23}^2-T_{12}^2 T_{33}$ which consists
of $F_4=3$ terms instead of $3!=6$ ones.
The number of terms still grows exponentially with $N$ but not super-exponential
like $N!$. Consequently, the computer-algebraic calculation of the characteristic polynomial of
$T$ is markedly faster than for a general $N\times N$-matrix.\\

We will write down the explicit formula for $\det T$ in order to substantiate our claim that the characteristic polynomial
of the general triangle can be given in closed form. To this end let $(d_1,\ldots,d_N)$ denote the diagonal of $T$ and
$(c_2,\ldots,c_N)$ its secondary diagonals. Further let ${\mathcal B}(m,k)$ denote the set of $(0,1)$-sequences of length
$m+k$ containing exactly $m$ zeroes and $k$ ones. There are $\binom{m+k}{k}$ such sequences. For any $\sigma\in{\mathcal B}(m,k)$
we define
\begin{equation}\label{A10}
n(i,\sigma)\equiv i+\sum_{j=1}^i \sigma(j),\quad i=1,\ldots,m+k
\;.
\end{equation}
Then
\begin{eqnarray}\label{A11}
\det\,T&=& \sum_{\footnotesize\begin{array}{c}m,k\in{\mathbb N}\\m+2k=N\end{array}}
(-1)^k \sum_{\sigma\in{\mathcal B}(m,k)}\prod_{i=1}^{m+k}\left\{
\begin{array}{l@{\quad : \quad}l}
d_{n(i,\sigma)}& \sigma(i)=0,\\
c_{n(i,\sigma)}^2 & \sigma(i)=1.
\end{array}
\right.
\end{eqnarray}

Another simplification occurs for the perturbation series of, say, a diagonal matrix $A$ with pairwise different eigenvalues
perturbed by a symmetric tridiagonal matrix $y\,B$. We will write the tridiagonal matrix $A+y\,B$ in the form

\begin{equation}\label{A2}
\left(
\begin{array}{ccccccccc}
 * & * & 0 & 0 & 0 & 0 & 0 & 0 & 0 \\
 * & * & * & 0 & 0 & 0 & 0 & 0 & 0 \\
 0 & * & * & y\sqrt{B_{n-2}} & 0 & 0 & 0 & 0 & 0 \\
 0 & 0 & y\sqrt{B_{n-2}} & A_{n-1}+{y\alpha}_{n-1} & y\sqrt{B_{n-1}} & 0 & 0 & 0 & 0 \\
 0 & 0 & 0 & y\sqrt{B_{n-1}} & A_n+{y\alpha}_n & y\sqrt{B_{n+1}} & 0 & 0 & 0 \\
 0 & 0 & 0 & 0 & y\sqrt{B_{n+1}} & A_{n+1}+{y\alpha}_{n+1} & * & 0 & 0 \\
 0 & 0 & 0 & 0 & 0 & * & * & * & 0 \\
 0 & 0 & 0 & 0 & 0 & 0 & * & * & * \\
 0 & 0 & 0 & 0 & 0 & 0 & 0 & * & *
\end{array}
\right),
\end{equation}
which is symmetric w.~r.~t.~the $n$-th diagonal entry (note that there is no $B_n$).
This symmetric form facilitates the notation in what follows. We introduce, for fixed $n$, the abbreviations
\begin{eqnarray}\label{A3a}
a_i&\equiv&A_i-A_n,\\
\label{A3b}
\beta_i&\equiv& \frac{\alpha_i-\alpha_n}{A_i-A_n},\quad(i\neq n)\\
\label{A3c}
b_i&\equiv& \frac{B_i}{A_i-A_n},\quad(i\neq n)
\;.
\end{eqnarray}
Let
\begin{equation}\label{A4}
x^{(n)}=A_n+\alpha_n y +\sum_{i=2}^\infty x_i^{(n)} y^i
\end{equation}
denote the perturbation series for the $n$-th eigenvalue $x^{(n)}$ of $A+yB$, where we have anticipated
the elementary result for the first two terms. Then the next terms are of the form
\begin{eqnarray}\label{A5a}
x_2^{(n)}&=&-b_{n-1}-b_{n+1}= -b_{n-1}+\mbox{RT},\\
\label{A5b}
x_3^{(n)}&=&b_{n-1}\beta_{n-1}+b_{n+1}\beta_{n+1}= b_{n-1}\beta_{n-1}+\mbox{RT},\\
\label{A5c}
x_4^{(n)}&=&-\frac{b_{n-2}b_{n-1}}{a_{n-1}}+\frac{b_{n-1}^2}{a_{n-1}}+\frac{b_{n-1}b_{n+1}}{a_{n-1}}-b_{n-1}\beta_{n-1}^2+\mbox{RT}
\;.
\end{eqnarray}
Here ``RT" (reflected terms) stands for the same terms as before but with reflected indices $n-i \leftrightarrow n+i$.
The above formulas also hold for small $n=1,2,3,\ldots$ or large $n=N,N-1,N-2,\ldots$ if the corresponding
$A_i$ and $B_i$ are set to zero. It is remarkable that the perturbation has a ``local" character in so far
as $x_{2k}^{(n)}$ and $x_{2k+1}^{(n)}$ only depend on matrix elements with indices $n-k,\ldots,n+k$.
Therefore it is possible to obtain the above results by computer-algebraic calculations with at most $5\times 5$- matrices.
We note that there are further simplifications for our case of
$A+y B =\op{\bi{s}}_1\cdot\op{\bi{s}}_2+y\,\op{\bi{s}}_2\cdot\op{\bi{s}}_3$, see (\ref{S16a})-(\ref{S16e}), which do not hold in the case
of general tridiagonal matrices.\\

A further property of tridiagonal matrices $T$ with non-vanishing entries in the secondary diagonals is that their eigenvalues
are never degenerate.\\
For the proof, let $\varphi$ be an eigenvector of $T$ with eigenvalue $x$, that is
\begin{equation}\label{A5}
A\,\varphi \equiv (T-x\; \Eins)\,\varphi=0
\;.
\end{equation}
It follows that $A_{11}\varphi_1+A_{12}\varphi_2=0$, or $\varphi_2=-\frac{A_{11}}{A_{12}}\varphi_1$ since $A_{12}=T_{12}\neq 0$.
Similarly, $A_{21}\varphi_1+A_{22}\varphi_2+A_{23}\varphi_3=0$ implies
\begin{equation}\label{A6}
\varphi_3=\frac{1}{A_{12}A_{23}}\left(A_{11}A_{22}-A_{12}A_{21} \right)\varphi_1
\;,
\end{equation}
since $A_{23}\neq 0$. By induction, one easily shows that
\begin{equation}\label{A7}
\varphi_n=\left(\prod_{i=1}^{n-1}A_{i,i+1} \right)^{-1}(-1)^{n-1}\,D^{(n-1)}\varphi_1,\quad n=2,\ldots,N
\;,
\end{equation}
where $D^{(n)}$ denotes the $n$-th principal minor of $A$, e.~g.~,
$D^{(1)}=A_{11}$, $D^{(2)}=A_{11}A_{22}-A_{12}A_{21}$, and so on.
(\ref{A7}) implies that the eigenspace of $T$ corresponding to the eigenvalue $x$ will be one-dimensional, which completes the proof.\\
As a by-product we have obtained an explicit formula for the eigenvector $\varphi$ corresponding
to the eigenvalue $x$ of a tridiagonal matrix $T$.

%%%%%%%%%%%%%%%%%%%%%%%%%%%%%%%%%%%%%%%%%%%%%%%%%%%%%%%%%%%%%%%%%%%%%%%%%%%%%%%%%%%%%%%%%%%%%%%%%%%%%%%%%%%%%%%%%%%%%%%%%%%%%%%%%%%%%%%%%%%%%%%%

\section*{Appendix B: Moments of the one-chain\label{sec:B}}

%%%%%%%%%%%%%%%%%%%%%%%%%%%%%%%%%%%%%%%%%%%%%%%%%%%%%%%%%%%%%%%%%%%%%%%%%%%%%%%%%%%%%%%%%%%%%%%%%%%%%%%%%%%%%%%%%%%%%%%%%%%%%%%%%%%%%%%%%%%%%%%%

Since the eigenvalues of the one-chain (or ``spin dimer") are\\
$e_S=\frac{1}{2}(S(S+1)-2 s(s+1))$ with degeneracy $2S+1$, its $k$-th moment is given by
\begin{equation}\label{B1}
p(k)=\frac{1}{(2s+1)^2}\sum_{S=0}^{2s}(2S+1)\left[\frac{1}{2}(S(S+1)-2 s(s+1))\right]^k
\;.
\end{equation}
This sum can be conveniently evaluated by computer-algebra software but it is not an explicit formula for the polynomials
$p(k)(r),\;r\equiv s(s+1)$ in the strict sense. However, such an explicit formula exists since (\ref{B1}) can be reduced to power sums,
which in turn are expressible by means of the Bernoulli numbers $B_n$:
\begin{equation}\label{B2a}
p(k)(r)=
\end{equation}
\begin{displaymath}
\frac{2^{-3 k-2}}{k+1} \sum_{\nu =0}^k (-1)^{\nu } (8 r+1)^{\nu }
   \binom{k+1}{\nu } \sum_{\mu =1}^{k-\nu +1} (4 r+1)^{\mu -1} \sum
   _{j=2 \mu }^{2 (k-\nu +1)} 2^j B_{j-2 \mu } \binom{j}{2 \mu }
   \binom{2 (k-\nu +1)}{j}\;.
\end{displaymath}

For example,
\begin{eqnarray}\label{B3a}
p(2)&=&\frac{1}{3}r^2\\
\label{B3b}
p(3)&=&-\frac{1}{6}r^2\\
\label{B3c}
p(4)&=&\frac{1}{15}r^2(2 - 2 r + 3 r^2)
\\
\label{B3d}
p(5)&=&-\frac{1}{6}r^2(2 - 2 r + 2 r^2)
\;.
\end{eqnarray}
From these polynomials we can, for example, calculate $p(1,2,3)$ by using (\ref{T10}):
\begin{eqnarray}\nonumber
p(1,2,3)&=&\frac{1}{8}\left\{-r^5 (5 + 2 r) + r \left[ 5 (-4 r p(3) + (-3 + 2 r) p(4)) - 12 p(5)\right]\right\}\\
\label{B4a}
&&\\
\label{B4b}
&=&\frac{1}{3}r^4 (1-2r)
\;,
\end{eqnarray}
in accordance with \cite{RBW} and \cite{SLR}.\\
Let us write $p(k)(r)=\sum_{n=2}^{k}a_n^{(k)} r^n$.
Then the first few leading coefficients of the polynomials $p(k)$
assume the following form:
\begin{eqnarray}
a_k^{(k)}&=&
\left\{
\begin{array}{rcl}
\frac{1}{k+1}&:&k\mbox{ even},\\
0&:& k\mbox{ odd},
\end{array}
\right.\\
a_{k-1}^{(k)}&=&
\left\{
\begin{array}{rcl}
\frac{k(2-k)}{12(k+1)}&:&k\mbox{ even},\\
\frac{1-k}{12}&:& k\mbox{ odd},
\end{array}
\right.\\
a_{k-2}^{(k)}&=&
\left\{
\begin{array}{rcl}
\frac{(k-2)^2 k (7 k+2)}{720 (k+1)}&:&k\mbox{ even},\\
\frac{(k-3) (k-1) (7 k-5)}{720} &:& k\mbox{ odd},
\end{array}
\right.\\
a_{k-3}^{(k)}&=&
\left\{
\begin{array}{rcl}
\frac{(4-k) (k-2) k \left(62 k^3-174 k^2+k+174\right)}{60480 (k+1)}
&:&k\mbox{ even},\\
\frac{(3-k) (k-1) \left(62 k^3-360 k^2+523 k-105\right)}{60480}&:& k\mbox{ odd}.
\end{array}
\right.
\end{eqnarray}

\begin{center}
\begin{figure}
\begin{center}
\includegraphics[width=12cm]{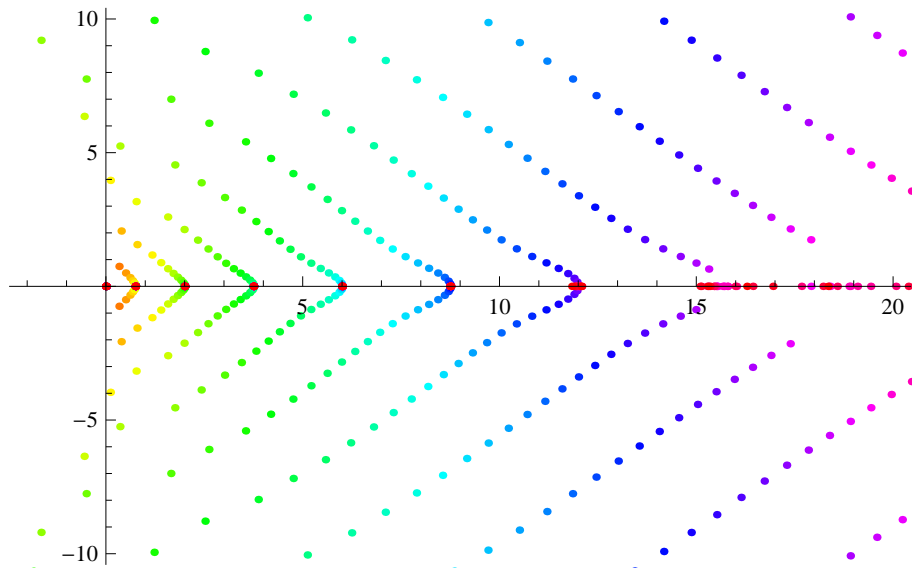}
\end{center}
\caption{\label{figG5}The sequence of complex zeroes of the dimer polynomials
$p(k),\;k=2,\ldots,50$, where different colors correspond to different values of $k$.
Obviously, besides the trivial value of $r=0$,
there are accumulation points for the physical values $r=s(s+1),\;s=1/2,1,3/2,2,5/2,3$.}
\end{figure}
\end{center}

Another representation of the polynomials $p(k)$ can be obtained from the
Euler-MacLaurin summation formula, see \cite{AS} 23.1.30. After some simplifications
we arrive at the following expression involving only a single summation:
\begin{eqnarray}\nonumber
p(k)&=&\frac{1}{(2 s+1)^2}\frac{1}{k+1} \\ \nonumber
&&
\left[
\sum_{m=1}^k \frac{2}{2^m} B_{2 m} \binom{k+1}{m} \left(\left(s^2\right)^{k-m+1} \,
   F\left(-m,-k+m-1;\frac{1}{2};\frac{16 s^2+8 s+1}{8 s^2}\right)\right.\right. \\  \nonumber
 &-& \left. (-s (s+1))^{k-m+1} \,
   F\left(-m,-k+m-1;\frac{1}{2};-\frac{1}{8 s (s+1)}\right)\right)\\  \label{B6}
&+& \left.\frac{1}{2}\left(s^{2 k}
 \left(4 k s+k+(2s+1)^2\right)+(-s (s+1))^k\,\left(k+(2 s+1)^2\right)
 \right)\right].
\end{eqnarray}
Note that the hypergeometric function $F(a,b;c;z)$ reduces to a polynomial in $z$ if
$a$ or $b$ are negative integers, see \cite{AS} 15.4.1. Hence (\ref{B6}) is a priori
a rational function of $s$. Only after the summation it turns out that the factor
$(2s+1)^2$ cancels and the remaining sum is a polynomial in $r=s(s+1)$.\\
The $p(k)$ can be viewed as polynomials in the complex variable $r$. Interestingly,
the physical values of $r=s(s+1),\; s=\frac{1}{2},1,\frac{3}{2},2,\ldots$ can be recovered
as accumulation points of the sequence of complex zeroes of all $p(k),\;k=2,3,4,\ldots$.
This is not rigorously proven but demonstrated by numerical evidence, see figure \ref{figG5}.

%%%%%%%%%%%%%%%%%%%%%%%%%%%%%%%%%%%%%%%%%%%%%%%%%%%%%%%%%%%%%%%%%%%%%%%%%%%%%%%%%%%%%%%%%%%%%%%%%%%%%%%%%%%%%%%%%%%%%%%%%%%%%%%%%%%%%%%%%%%%%%%

%%%%%%%%%%%%%%%%%%%%%%%%%%%%%%%%%%%%%%%%%%%%%%%%%%%%%%%%%%%%%%%%%%%%%%%%%%%%%%%%%%%%%%%%%%%%%%%%%%%%%%%%%%%%%%%%%%%%%%%%%%%%%%%%%%%%%%%%%%%%%%%%%%
\end{document}